# Design Strategies and Knowledge in Object-Oriented Programming: Effects of Experience

**Françoise Détienne***

Projet de Psychologie Ergonomique

INRIA

Domaine de Voluceau, Rocquencourt, BP 105, 78153 Le Chesnay, cedex,

France.

running head title: Design Strategies and knowledge in OOP

* Françoise Détienne is a cognitive psychologist with an interest in software design; she is a researcher in the Ergonomic Psychology group of INRIA.

## ABSTRACT

An empirical study was conducted to analyse design strategies and knowledge used in object-oriented software design. Eight professional programmers experienced with procedural programming languages and either experienced or not experienced in object-oriented programming participated in this experiment. They were asked to design a program for a procedural problem and a declarative problem.

We have concentrated our analysis on the design strategies related to two central aspects of the object-oriented paradigm: (1) associating actions, i.e., execution steps, of a complex plan to different objects and revising a complex plan, and (2) defining simple plans at different levels in the class hierarchy.

As regards the development of complex plans elements attached to different objects, our results show that, for beginners in OOP, the description of objects and the description of actions are not always integrated in an early design phase, particularly for the declarative problem whereas, for the programmers experienced in OOP, the description of objects and the description of actions tend to be integrated in their first drafts of solutions whichever the problem type.

Most of the first drafts of solutions were structured around the objects, whatever the experience of subjects in OOP. However, by analysing the order in which actions are generated, we found that, for the programmers experienced in OOP, methods were grouped together mainly by membership to the same class whereas, for beginners in OOP, methods were grouped together mainly by functional similarity and execution order. Furthermore, with only a little experience in OOP complex plans are revised more often.

As concerns the development of simple plans at different levels in the class hierarchy, our results indicate that with a little OOP experience, simple plans are developed either in a top-down manner or in a bottom-up manner. With more OOP experience, our results show that a simple plan is developed in a strictly top-down manner.

The analysis of design strategies reveal the use of different knowledge according to subjects' language experience: (1) schemas related to procedural languages; actions are organized in an execution order, or (2) schemas related to object-oriented languages; actions and objects are integrated, and actions are organised around objects.

# CONTENTS

**APPENDIX C: Characteristics of the programs produced by the subjects**

**APPENDIX C1: Characteristics of the programs produced for the procedural problem**

**APPENDIX C2: Characteristics of the programs produced for the declarative problem**

**APPENDIX C3: Characteristics of the programs: average by subject type**

1. INTRODUCTION

The topic of this research is the activity of object-oriented software design. An empirical study was conducted to analyse design strategies used by programmers experienced with procedural programming languages and programmers either experienced in object-oriented programming (OOP) or not experienced in OOP. We will analyse the shift of strategies when experience in OOP increases. The study of strategy shifts should make it possible to analyse the transfer of knowledge from procedural languages and to analyse the knowledge gained through experience with the object-oriented programming language.

The originality of this study is twofold. First, to date, there have been very few empirical studies on object-oriented programming[1] (Kim & Lerch, 1992; Lange & Moher, 1989). Most empirical studies on software design have been conducted with procedural, functional or declarative languages. Whereas design strategies and knowledge used with these kinds of languages have been extensively analysed, particularly for procedural languages, these issues have not yet been addressed, as far as we know, for object-oriented languages.

Second, insofar as new programming languages are becoming increasingly available to experienced programmers, the role of previous knowledge in the design activity of experts using new languages seems an important psychological issue to address. Object-oriented languages are often used by programmers having previous experience with procedural languages. The role of previous knowledge is important when using a new language. This issue has been addressed only in a few studies (Scholtz & Wiedenbeck, 1990, 1992), and these studies were conducted with languages from the same paradigm.

The originality of the present study lies in the fact that it deals with this issue for languages from different paradigms, i.e., procedural and object-oriented paradigms.

Our study will examine design strategies related to two central aspects of the object-oriented paradigm: (1) associating actions, i.e. execution steps, of a complex plan to different objects and revising a complex plan, and (2) defining simple plans at different levels in the class hierarchy. We expect different design strategies according to the programmers experience with OOP.

In the next section, we present a brief review of the literature on questions related to the knowledge and design strategies used in the design activity with procedural languages. Then, we discuss the possible characteristics of a program's representation and knowledge used in the design activity with object-oriented languages. This enables us to formulate hypotheses that are tested in the empirical study. They concern the design strategies used with an object-oriented language and the role of knowledge constructed with this kind of language as well as the role of knowledge constructed through previous experience with procedural languages.

**1.1. Knowledge and Design Strategies Used with Procedural Languages**

In the schema-based theoretical framework, evolving from novice to expert is assumed to involve the construction of knowledge structures such as schemas of programming (Détienne, 1990a, 1990b; Rist, 1986; Robertson & Yu, 1990; Soloway & Ehrlich, 1984). A plan represents portions of code which achieve a common goal and a program is viewed as a set of plans, complex and basic plans, which are merged together to achieve the problem goal (Rist, 1991). When a plan is memorised as a knowledge structure, it is called a plan schema (Rist, 1986) or programming plan (Soloway, Ehrlich & Bonar, 84) or schema (Détienne, 1990a). In this paper, the term "plan" will refer to representations of solutions constructed to achieve given goals in the design activity, whereas the term "schema" will refer to the knowledge structure, stored in memory, which may be evoked during design for elaborating a plan. A schema is a data structure which represents generic concepts stored in memory. In the procedural programming domain, a schema may be described as a set of actions, i.e., execution steps, with some constraints on the order of execution of actions. Knowledge constructed with procedural languages, such as Pascal, have been described with this concept (Soloway & Ehrlich, 1982).

Experts are assumed to possess schemas representing information on problems, such as the possible data structure and the possible functions in a particular problem domain[2], as well as schemas dependent on the programming domain. Détienne (1990b) provides empirical support for this hypothesis on the basis of data collected in a program understanding task and a program recall task. It is shown that schemas are activated during program reading; schemas may be activated in a bottom-up manner by cues of the code and in a top-down manner when activated schemas evoke other schemas they are linked to. These activations create expectations about what functions are performed by the program and how they are performed. Other empirical evidence from the litterature is discussed by Détienne (1990a).

Among the schemas relative to the programming domain, an important distinction is made between "tactical plans" or "strategic plans" (Soloway et al., 1982), which represent more or less general solutions independent of a particular language, and "implementation plans" which are language-dependent.

Selection rules are also assumed to be constructed through expertise in programming (Black , Kay & Soloway, 1986). Even though it was not studied in the programming domain, Kay and Black's study (1985) on learning to use a text editing system brings evidence on the use of selection rules when experience increases. They found that experts combined simple plans into more compound plans to accomplish major goals and developed rules for selecting the best plan to achieve a given goal in a given situation. In their study, there were often several plans that could be instantiated to achieve a given goal. To be sure that the correct plan was chosen from the set of applicable plans, the links that connected these plans to the goal were conditions or selection rules that had to be met before a given plan was chosen.

Some studies (Adelson & Soloway, 1985; Davies, 1991; Guindon, 1990; Rist, 1991) show the role of knowledge, such as schemas, in procedural software design. Design strategies tend to vary according to the programmer's expertise. In Adelson and Soloway's study , it was found that experienced programmers tend to use top-down and breadth-first strategies when they solve familiar problems.

Design strategies accounting for the order in which parts of plans are developed by the subjects may reveal the use of schemas. If a schema is evoked, actions of this schema would be made available in their schema or execution order; in this case, in notes written by subjects, the actions composing plans should appear in their execution order. In Rist's study, the design activity of experienced and novice programmers was compared. It was shown that when the programmers know all the required abstract and detailed schemas, then the plans are developed in a top-down and forward manner, i.e., from the first to the last executable action. Evidence for schema retrieval consisted of actions appearing in their executable or schema order. When a schema cannot be retrieved, it was shown that the plans are developed in a bottom-up and backward manner, i.e., from the last to the first executable action.

**1.2. Program Representation and Knowledge Constructed with Object-Oriented Languages**

There is an important difference between the object-oriented paradigm and the procedural paradigm. With a procedural paradigm, data and functions are separated whereas, with an object-oriented paradigm they are integrated. Objects are program entities which integrate a structure defined by a type as well as functionalities. Objects are usually instances of classes. A class is defined as a structure (a type) and a set of methods. A method is a function attached to a class that describes a part of the behaviour of the objects which are instances of this class.

According to Rist (Rist, 1994; Rist & Terwilliger, 1994), plans and objects are orthogonal. Plan elements may be attached to different classes/objects and the links should be made explicit. One plan can use many objects and one object can take part in many plans. This reflects the real world, where a plan can use many objects (the plan for a cake uses flour, eggs, water, and so on), and an object can be used in many plans (an egg can be whipped, fried, boiled, and so on). Rist (1994) illustrates this characteristic through the following example.

"Consider an OO system that models an automatic teller machine (ATM) in a bank. The main objects in the system would be a BANK, that contains a set of CUSTOMERs. A CUSTOMER has a set of ACCOUNTs.... The account would have a MENU as a system interface, so that the user could input commands and queries.... Assume that a user wishes to withdraw money from her account. To do this, she should first enter her customer identifier into the ATM.... The BANK would validate it, and use it to look up the particular CUSTOMER. The CUSTOMER module would then ask for a password... and transfer control to the MENU. The MENU would ask for the type of account, and the action on that account; assume that the user chooses the cheque account, and enters 'W' to withdraw money. The MENU system would validate the input, then call a routine in the [ACCOUNT] to actually change the account balance. Note that the code that actually implements the goal is a small section of code in the ACCOUNT class.... The ... plan has fragments of code in each of the other classes."

We assume that a simple plan, representing a part of a more complex plan, may be developed at different levels in the class hierarchy. A simple plan may be defined at the level of a class and redefined at the level of its subclasses. A typical example is an initialisation method which may be defined at the superclass level and at the subclasses levels. In some cases, using the inheritance property, the plan should be developed only at the upper level; there is no need to develop it at subclasses level because it is inherited by them and there are no specificities to be added at the lower level.

To summarise, the parts of a "complex plan" achieving the problem's main goal(s) may be attached to various classes in an OO program. A single part which is attached to one class is referred to as "simple plan": this simple plan may be implemented at different levels in the hierarchical structure of classes.

It is unclear what the characteristics of schemas related to OOP languages are. In a general way, we will assume that such schemas would integrate objects and procedures which use these objects to achieve goals. According to the prominence of the procedure's characteristics or the object's characteristics, two alternative views of schemas are possible.

One view is to assume that a schema is organised around one main procedure to achieve a goal. In this case, it would integrate characteristics of the procedure and characteristics of objects used by this procedure. Some characteristics of the procedure would be similar to characteristics of schemas related to procedural languages: actions to achieve a goal, some actions being more important than others. However, this kind of schema would be different from a schema related to procedural languages in as much as it would also integrate objects. It may represent the links between actions and objects, different actions being linked to different objects. This schema would integrate all actions common to a single plan.

An alternative view is to assume that a schema is organised around one main object. In this case, it would integrate the characteristics of this object and, in addition, actions linked to this object. This schema would integrate actions related to several plans but linked to the same object.

Design methodologies (Booch, 1991) encourage the programmer to focus on objects first rather than on a functional decomposition in the construction of a solution. Proponents of OO have assumed that the knowledge in the problem domain should help the subjects to carry out such a decomposition (Rosson & Alpert, 1988). It should be remarked that our approach is quite different but not incompatible with Rosson and Alpert's approach. We assume that the use of knowledge structures constructed with OOP, such as schemas abstracted when developing plans, should facilitate such a solution decomposition.

**1.3. Shifting from Procedural to Object-Oriented Languages**

We have conducted an empirical study on the design activity with an object-oriented language, the $CO_2$ language, and an object-oriented data base management system, the $O_2$ system (Lecluse & Richard, 1989; 0.Deux et al., 1989). Two types of programmers participated in the experiment. All were experienced with procedural languages; half of them were also experienced in OOP while the other half were beginners in OOP. One question addressed concerns the characteristics of design strategies depending on the programmers' knowledge related to OOP languages.

This article focuses on the design strategies used in OOP and the use of knowledge. We have concentrated our analysis on the design mechanisms related to two central aspects of the object-oriented paradigm: the location of the code for a complex plan in different classes and the definition of simple plans at different levels in the class hierarchy. We have also analysed the effects of problem characteristics on the design activity.

**Developing a Complex Plan Attached to Different Objects**

The first type of analysis concerns the strategies used for developing elements of complex plans attached to different objects in the early design activity. We will analyse the order in which the parts of complex plans are developed and attached to various classes in an OO program. We consider that the order in which parts of plans are developed reflects the characteristics of the knowledge used by programmers.

For programmers who are experienced in OOP, we expect complex plans to be developed in a breadth-first and top-down manner. The use of knowledge structures, such as schemas related to OO languages grouping actions and objects together, should allow subjects (1) to develop a plan, at a high level of abstraction first and in a balanced way, before developing less abstract levels, and (2) to integrate the description of actions and the description of objects in their first drafts of the solutions. Furthermore, one question is whether or not complex plans are developed on the basis of the objects representation, i.e., objects are developed first and then actions are developed in an order related to the association to a single class. This would reflect the use of schemas which are organised around main objects.

In an OO system, the objects are explicit and are used to structure the system, so design based around the objects should be a common strategy. In a procedural language, the action are primary, and any objects are implicit in the plan. From this perpective, the normal design way is to pursue the links between actions, and see what objects appear on the way. For programmers who are beginners in OOP but experienced in procedural languages, we expect the description of actions and the description of objects to be separate in the first drafts of the solutions produced by these subjects. We expect complex plans to be developed on the basis of representation of actions, i.e., actions are developed first then objects are associated to actions. Furthermore, we expect actions which are components of complex plans to be developed in their execution order. The use of knowledge structures, such as schemas related to procedural languages, should lead subjects to develop plans from representation of actions; these schemas represent actions and information on their execution order but lack information on the relationship between plans actions and objects. The difficulties then encountered in associating elements of complex plans to objects should preclude a strictly top-down and breadth-first way of developing plans.

So the difference in the order in which plans are developed should reflect the use of different kinds of knowledge, either schemas related to OO languages in which actions are organised around objects or schemas related to procedural languages in which actions are represented in their execution order. In so far as the use of knowledge, such as schemas related to procedural languages, is not appropriate, we should expect there to be more errors and also more revisions of the decomposition of classes by beginners in OOP than by programmers experienced in OOP.

**Developing Simple Plans at Different Levels in the Class Hierarchy**

A routine that is used in several places can be coded in each of these places, or coded in a parent class and inherited. The second type of analysis concerns the strategies used for developing simple plans at different levels in the hierarchically-organised structure of classes. We will analyse the order in which functionally similar simple plans are implemented at different levels in the structure of classes. We expect these plans to be developed at a general level first, i.e., in a top-down manner, by programmers experienced in OOP and at a specific level first, i.e., in a bottom-up manner, by beginners in OOP.

**Effect of Problem Characteristics**

Previous studies have shown effects of problem characteristics on design strategies. Two kinds of design problems have been used in this experiment. One problem is "declarative"; it has been shown by Hoc, in a previous study (Hoc, 1983, 1988), that the data structure guides the program development. The other problem is "procedural"; the structure of the procedure has been shown to guide the program development. In the paradigm of object-oriented programming, identifying objects and their characteristics is important in the design process. In a declarative problem, these aspects may be more straightforward to analyse. So an object-oriented language could make the design activity easier for a declarative problem than for a procedural problem, at least for the beginners in OOP. This will be measured by the strategies used and the errors produced according to the problem type.

In summary, we expect:

-hypothesis 1a: programmers who are experienced in OOP to make explicit action and object links in their first drafts of the solutions and programmers who are beginners in OOP not to link actions and objects in their first drafts of the solutions

-hypothesis1b: complex plans to be developed on the basis of representation of actions by programmers who are beginners in OOP but experienced in procedural languages and on the basis of representation of objects by programmers who are experienced in OOP

-hypothesis1c: actions of complex plans to be developed more frequently in their execution order by programmers who are beginners in OOP but experienced in procedural languages than by programmers who are experienced in OOP

-hypothesis2: complex plans to be developed in a more breadth-first manner by programmers who are experienced in OOP and in a more-depth first manner by programmers who are beginners in OOP

-hypothesis3a: more revisions of the decomposition of classes by programmers who are beginners in OOP than by programmers who are experienced in OOP

-hypothesis3b: more errors in the decomposition of classes by programmers who are beginners in OOP than by programmers who are experienced in OOP

-hypothesis4: simple plans to be developed in a top-down manner by programmers who are experienced in OOP and in a bottom-up manner by programmers who are beginners in OOP

-hypothesis5: the design activity to be easier for a declarative problem than for a procedural problem, at least for programmers who are beginners in OOP

In section 2, the methodology of our experiment is presented. Results are presented in section 3, and discussed in section 4.

**2. THE $CO_2$ EXPERIMENT**

**2.1. Subjects**

Eight professional programmers participated in this experiment. All had several years of programming practice with classical procedural languages such as C, Pascal, Basic and Fortran. Four of them were "beginners in OOP"; they will be referred to as B1, B2, B3, and B4 in the result sections. They had no practical experience with the OOP paradigm. The four others, "experienced in OOP", had practical experience with $CO_2$, the object-oriented language under study; they will be referred to as E1, E2, E3, and E4 in the result sections. E1 also knew Smalltalk and E3 knew Booch's method. The subjects' participation was funded by the GIP Altaïr. The beginners in OOP were paid for their three-day participation and, the programmers experienced in OOP took part in the experiment as a part of their current work. A more detailed profile of subjects is presented in Appendix A.

**2.2 Procedure**

Each subject had two problems to program with the $CO_2$ system. The material consisted of two problems of management:

-a library management problem which was a slightly modified version of a problem classified in a previous experiment (Hoc, 1983; 1988) as a procedural problem,

-a financial management problem which was a slightly modified version of a problem classified as a declarative problem.

Problem statements are given in Appendix B. The task domain was familiar to the subjects. As the second problem was not adequate for encouraging the use of inheritance, we added a remark to encourage the subjects to use a predefined class. Using this predefined class correctly would mean to use it as a superclass for one of the particular solution's classes.

The order of problem presentation was counter-balanced. The programmers not experienced in object-oriented programming had one day to program each problem, whereas the programmers experienced in object-oriented programming had half a day. This proved to be sufficient to develop the program at least as far as the beginners were capable of doing. Some characteristics of the programs produced by the subjects (number of defined classes, number of defined methods, number of intructions, etc.) are given in Appendix C.

Previous to the phase of program design, the four beginners in OOP received a one-day theoretical course in OOP as well as in the $CO_2$ language , and some training with the $CO_2$ environment.

The subjects were asked to verbalise while designing their programs. They were allowed to ask questions to several experts in OOP whenever they had problems they were not able to overcome. All the subjects had at their disposal a manual for the system, a theoretical paper on object-oriented programming and an example-program written in $CO_2$, solving a problem different from the experimental ones, and its problem statements. After the programming phase, the subjects had to answer questions on the difficulties they had experienced during the experiment.

Subjects were allowed to use the $CO_2$ environment in addition to paper and pencil. This provided them with the opportunity to compile their programs, to run them on test data, and to access and reuse previous programs. Our subjects, who were experienced programmers, did not exhibit any problems in using the programming environment, so this did not seem to increase the complexity of the task. Furthermore, this provided them with a means of evaluating their program[3].

We collected the subjects' verbalisations, successive versions of programs under development, notes written during the realisation of the task, and questions asked to the experts. The main data was the order in which the different traces of the activity were made, i.e., the order for writing notes and coding programs with the verbalisation recorded simultaneously. We collected a total of about seventy hours of recorded verbal protocols.

The final versions of programs were given for evaluation to two experts in object-oriented programming. They were asked to detect and report errors as well as "inelegances" in design and style. They had to rank the reported errors and inelegances by order of relative importance, to classify them and to make explicit their criteria of classification.

**2.3. The CO$_2$ Program Structure**

The O$_2$ system is an object-oriented data base system. A "classical" language is used mainly to write the methods. In the version of the system used for our experiment, this language is the CO$_2$ language. It is a modified version of C: an object-oriented layer, the O$_2$ language, is added to the C language.

A program is composed of two parts:

-a *declarative part* in which are defined computational entities and the relations between entities. This part is called the "class model" (or "schema" in the terminology of O$_2$ system designers) and is written in O$_2$. It consists of the names of classes, plus the signatures (name, type, and argument) of the attributes and methods.

-a *procedural part* which consists of the bodies of methods. This is written in CO$_2$.

The following excerpt illustrates the declarative part of a program, in Figure 1a, and the procedural part of the same program, in Figure 1b.

# Figure 1a and Figure 1b about here

In this example, the classes Proceedings and Journal are subclasses of the class Book. Thus they inherit its structure. This means that the type of the class Proceedings is a tuple with three fields: title, year and place. The subclasses also inherit the functionalities of their superclass. This means that the methods "title" and "show" are inherited by the class Proceedings. In this example, the method "show" is redefined in the two subclasses.

**3. DESIGN STRATEGIES**

We observed that, whatever their experience in OOP, subjects tried to define the class model before they implemented the methods in detail. This behaviour is driven by specific characteristics of the $O_2$ system because it is not possible to use an object in a method body if the object has not been completely specified before. The plan programmers try to follow is hierarchical. They try to develop the most abstract aspects of the solution, i.e., entities of the declarative part, before writing the code of methods which is a refinement of some functional aspects defined in the declarative part.

Whereas the global activity of all subjects tends to be organised in a hierarchical manner, the design strategies they follow tend to differ according to their experience in OOP. We have concentrated our analysis[4] on the activities related to two central aspects of the object-oriented paradigm: the association of actions to different objects and the development of similar methods at different hierarchical levels. The first type of analysis concerns the strategies used for developing a complex plan attached to different objects in the early phase of design when subjects construct the class model. The second type of analysis concerns the strategies used for developing simple plans at different levels in the class hierarchy. We will analyse the strategies used in an early phase of design when the subjects first develop the class model, and the strategies used, later on, when reusing code for writing the code of similar program units. The third type of analysis concerns the extent to which the complex plans developed in an early design phase are revised by subjects later on.

**3.1 Developing a Complex Plan Attached to Different Objects**

We have analysed the first drafts of the solutions produced by subjects before they start writing the code of methods, including the order in which the subjects produce the entities represented in these drafts. This provides information on the strategies followed for developing complex plans to achieve the main goals of the problem; actions which are parts of these complex plans are attached to different objects.

When first elaborating the class model, the entities evoked and used by programmers are objects, actions and different types of links between them. When describing the class model in terms of the programming language, objects should be mapped to computational entities such as the static description of classes, i.e., the class name, its type and attributes. Actions should be mapped to computational entities such as methods definition or parts of methods, and links should be mapped to computational entities such as inheritance links and composition links between classes and associations between classes and methods.

Four complementary analyses have been conducted. A first analysis consisted in identifying whether or not the description of actions and the description of objects are integrated in the first drafts of the solutions that the subjects produced.

Second, we analysed the characteristics of the strategy used in elaborating the links between objects and actions. Two strategies are distinguished: a declaratively-driven strategy and a procedurally-driven strategy.

Third, when the subjects defined several methods consecutively in the class model, we analysed what kind of relationship exists between these methods. Three relationships were distinguished: membership to the same class, functional similarity and execution order.

Fourth, we analysed whether the subjects develop methods in a breadth-first manner, i.e., all methods are defined in the class model before being refined when writing the body code, or in a depth-first manner, i.e., a method is defined in the class model then refined before another method is defined then refined and so on.

**Analysis1: Integration versus No Integration between the Description of Actions and the Description of Objects**

An analysis of the early phase of design has been conducted in order to describe how objects, actions and links between them are elaborated and how they are mapped with computational entities. Figure 2a and Figure 2b present some characteristics of the first drafts of solution produced by subjects.

Shifting from one draft to another corresponds to keeping some of the same entities but reorganising them and/or introducing new entities or, to elaborating new types (action versus objects) of entities without explicit relationships with the entities defined previously. Thus a draft does not correspond necessarily to the notes taken on a single sheet of paper; it may correspond to one part of a sheet of paper, to one sheet or to several sheets of paper.

For example, a programmer experienced in OOP, "E1", for the procedural problem, starts by taking notes to represent a list of objects which should be identified in the program as distinct classes ("1. list of objects as classes"). Then he writes a second draft with another list of objects to be defined as classes which is a revised version of the list written before ("2. list of objects as classes"). In a third draft, he writes a list of actions that should correspond to distinct methods. For each action, he tries to identify one or several objects to which this action should be associated. At the same time, this subject conducts a mental simulation with some of these actions ("3. list of actions; for each action, several objects are associated + simulation").

## Figure 2a and Figure 2b about here

We consider that there is no integration between the description of objects and the description of actions when we observe either "t0. list of objects (as classes), t1. list of actions (as methods)..." or "t0. list of actions (as methods), t1. list of objects (as classes)...", even if at t1, for each action (or object) of the list, the subject tries to associate it to an object (or action). We consider that there is integration between the description of objects and the description of actions when we observe that there is not such a separation between the representation of objects and the representation of actions. Results by subject type are presented in Figure 3a for the procedural problem and Figure 3b for the declarative problem.

## Figure 3a and Figure 3b about here

For the declarative problem, the strategies followed by subjects tend to be different according to their experience in OOP (Fisher exact probability test: p=.07). This is not the case for the procedural problem. For the declarative problem, the beginners tend define the actions and objects separately, whereas the programmers experienced in OOP tend to integrate both. Our results support hypothesis 1a for the declarative problem but not for the procedural problem.

Whatever the problem type, in four observations, beginners in OOP tend to describe objects and actions separately while in the four other observations, they integrate the description of actions and the description of objects. An example of no integration is as follows. A beginner takes notes such as those shown in Figure 4.

## Figure 4 about here

In this subject's notes, there is a clear visual separation between the description of objects in the upper part and the description of actions in the lower part, this separation being emphasised by a line drawn by the subject.

In seven of eight observations, the description of objects and the description of actions are integrated for the programmers experienced in OOP. Only one experienced subject (E1 for procedural problem) created an intermediate representation consisting of a separate description of objects and actions. He takes notes such as those shown in Figure 5.

## Figure 5 about here

In this subject's notes, objects are listed below the title "definition of data" and actions are listed below the title "data processing". The subject makes no explicit links between objects and actions.

Our results tend to show that beginners in OOP describe objects and actions separately in their first draft of the solution more often than programmers experienced in OOP do, in particular for the declarative problem. This is consistent with our assumption that programmers who are experienced in OOP use knowledge in which actions and objects are integrated. In contrast, this knowledge would be less available for beginners in OOP. Furthermore, in their knowledge related to procedural programming languages, actions representing parts of complex plans and objects would be separate. This would explain why integration between objects and actions was less often observed in these subjects' first drafts of a solution.

**Analysis 2: Declaratively-Driven Strategy versus Procedurally-Driven Strategy**

Using a declaratively-driven strategy, the subject produces a draft of the solution made up of a list of objects (classes) with actions associated to objects. This is described in Figure 4 by "list of objects as classes; for each object, several actions are associated".

With this strategy, the solution is structured around objects and the actions are associated to the objects.

Using a procedurally-driven strategy, the subject produces a draft of the solution made up of a list of actions (methods) with objects associated to actions. This is described in Figure 2a and Figure 2b by "list of actions as methods; for each action, several objects are associated." With this strategy, the solution is structured around actions and the objects are associated to the actions. Results by subject type and by problem type are presented in Figure 6.

## Figure6 about here

Statistically, we did not find any significant difference according to the problem type or according to the subjects' experience in OOP.

Programmers experienced in OOP follow the declaratively-driven strategy in seven out of eight observations. Objects, actions and associations between them are defined in an iterative manner. E4, for example, defined the first draft of a solution for the procedural problem as a list of classes. In the next draft, there is a more complete list of classes with methods associated to classes. In as much as the structure of the solution is organised around objects, we have considered the strategy to be declaratively-driven. This draft is presented in Figure 7.

## Figure 7 about here

When writing this class model, we observed that this subject first lists the classes then adds the methods associated to classes. It should be remarked that the methods "enter", "print" and "match" are associated only to the super class. We will see in a further analysis presented in section 3.2. that the strategy for defining the general entities first versus the specific entities first varies according to the subject's experience with OOP. This example also illustrates a typical strategy of experienced subjects which involves defining only the general entities.

Beginners in OOP follow the procedurally-driven strategy in three out of nine observations and follow the declaratively-driven strategy in six out of nine observations. Objects, actions and associations between them are defined in an iterative manner. An example of the procedurally-driven strategy is as follows. The subject "B 1" for the procedural problem takes notes like those shown in Figure 8.

## Figure 8 about here

We have here a list of methods which the subject tries to associate to classes. In these notes and in the verbal protocol, he makes explicit the association for the method Display only. A link from the other methods to their classes was postponed.

For one beginner in OOP, a shift of strategies was observed. The subject "B1", for the procedural problem, follows first a procedurally-driven strategy, then a declaratively-driven strategy and then a procedurally-driven strategy again. It should be remarked that this shift of strategy explains why the total of strategies followed by beginners in Figure 6 is nine and not eight.

Whatever the strategy followed by beginners in OOP, we observed that many questions were raised about the associations between actions and objects. The choice of the association between objects and actions is sometimes postponed until the elaboration of the procedural part of their programs.

Our results do not support hypothesis 1b: complex plans to be developed on the basis of actions representation by programmers who are beginners in OOP but experienced in procedural languages and on the basis of objects representation by programmers who are experienced in OOP. The procedurally-driven strategy is not used significantly more often by beginners in OOP than by programmers experienced in OOP. Our results do not show any effect of the type of problems on the choice of strategy.

The use of the declaratively-driven strategy may reveal, for programmers experienced in OOP, the use of schemas related to OO languages in which objects and actions are integrated and in which actions are organized around objects. However, an interpretation of the use of this declaratively-driven strategy by the beginners in OOP as well as by the programmers experienced in OOP could be that, whatever their knowledge in OOP, subjects try to construct their first drafts of a solution with a structure which conforms to the structure of OOP programs. Indeed, in the class model, a list of classes should be defined with, for each class, a list of actions associated to it in order for the class model to be syntactically correct.

**Analysis 3: Relationships between Methods Defined in a Row**

Even if most of the first drafts of a solution are structured in a declarative way, the order in which actions are developed may vary according to the subjects' experience in OOP. So, we have conducted a complementary analysis which concerns the order in which methods are defined in the class model. When first elaborating the class model, the subjects define lists of methods. We have analysed what kind of relationships exists between methods defined in a row, i.e., consecutively. Three relationships have been distinguished:

-(1) membership in the same class: the subject produces in a row the definition of several methods/actions associated explicitly to the same class. For example: initialisation, search, display associated to the class book.

-(2) functional similarity: the subject produces in a row the definition of several methods/actions which are functionally similar. For example: initialisation of books, initialisation of proceedings and, initialisation of journal.

-(3) execution order: the subject produces in a row the definition of several methods/actions in their execution order. For example: enter the day of the flight, calculate the consumption and update member's information. This may be done with simulation.

Units of coding, i.e., chunks of methods, have been defined on the basis of change of category. As long as the relationship between methods defined in a row was one of these three categories, we have considered these methods as belonging to a single unit. On very rare occasions, two categories could have explained the relationship between methods defined in a row, either (1) and (2) or (1) and (3). In these cases, we decided to use the category (1) "membership in the same class" as it corresponded to the weakest hypothesis conforming with the semantics of OOP languages.

Figure 9 presents the number and percentage of chunks of methods produced by subjects according to the relationships between methods of a chunk.

## Figure 9 about here

Coding a set of methods within the same class as a single chunk ("membership in the same class) is significantly more frequent for the programmers experienced in OOP than for beginners in OOP (Mann-Whitney U test: P=.002). Chunks based on the execution order are significantly more frequent for beginners in OOP than for the programmers experienced in OOP (Mann-Whitney U test: P=.001). This result support hypothesis1c: actions of complex plans to be developed more frequently in their execution order by programmers who are beginners in OOP but experienced in procedural languages than by programmers who are experienced in OOP.

There is no significant difference between the groups in the use of the relationship "functional similarity". There is no significant effect of the problem type on the relationship guiding methods definition.

A striking result is that the main relationship between methods defined as a chunk by programmers experienced in OOP is membership in the same class. Defining methods in the execution order is very rare. The picture for beginners in OOP is quite different. The most frequent relationships between methods defined as a chunk are functional similarity followed by execution order. The least frequent relationship is membership in the same class. For methods defined in their execution order, this is often done by mental simulation: in one case for programmers experienced in OOP, in four cases for beginners in OOP.

For the programmers experienced in OOP, our result reflects the use of knowledge in which objects and actions are integrated and, in which object representation is prevalent. This could explain why the main relationship between methods defined as a chunk by these subjects is membership in the same class. For the beginners in OOP, our result reflects the use of knowledge related to procedural languages in which representation of actions and information on the execution order of actions are prevalent. This could explain why the main relationship between methods defined as a chunk are functional similarity and execution order.

**Analysis 4: Breadth-First versus Depth-First Strategy**

When a solution is elaborated in a breadth-first manner, the complete solution is elaborated at a given level of abstraction before being developed at the next level of abstraction. In terms of methods defined in the class model, the solution would be elaborated in a breadth-first manner if most methods were simply named and listed before any code body was written. A strictly depth-first strategy would consist of defining a method and writing the code of its body before defining another method. Figure 10 presents the numbers of methods defined in the class model before the elaboration of the procedural part. These numbers are given for both kinds of subjects and both kinds of problem.

Figure 10 about here

The numbers shown in Figure 10 do not reveal purely depth-first strategies (none of the numbers is one). However, we consider that low numbers reveal a rather depth-first strategy and high numbers reveal a more breadth-first strategy.

The number of methods defined by the beginners in OOP tends to be lower than the number of methods defined by programmers experienced in OOP. However this difference is not statistically significant. The problem type has no significant effect. This result does not support hypothesis2: complex plans to be developed in a more breadth-first manner by programmers who are experienced in OOP and in a more-depth first manner by programmers who are beginners in OOP.

It should be remarked that we have based our analysis on the absolute numbers of methods defined by subject rather that on the percentage of methods defined before developing the procedural part compared to the total of methods defined/coded at the end of the experiment by each subject. Our working hypothesis is that the total numbers which would have been defined/coded in a complete solution would have been the same for all subjects if they, in particular the beginners, had the time to develop a complete solution. However, an alternative interpretation of our data could be that the smaller number of methods defined first by beginners could be due to the smaller number of methods coded by these subjects (as shown in Appendix C3).

For the programmers following a depth-first strategy, we have analysed what kinds of methods were elaborated first, at an abstract level. Methods developed first correspond to the main actions or goals to solve: for example, the search method for the management problem.

**3.2. Developing Simple Plans at Different Levels in the Class Hierarchy**

With the OOP paradigm, parts of a complex plan (methods) can be developed at various levels in the class inheritance hierarchy. We observed that: (1) inheritance of structural properties was difficult to use for the beginners in OOP and, (2) inheritance of functional properties was difficult for both the beginners in OOP and some of the programmers experienced in OOP.

We have conducted two complementary analyses to describe the strategies used by subjects in elaborating simple plans at different levels in the class hierarchy.

First, we analysed cases in which similar methods, i.e., methods performing the same function (e.g. initialisation) but not necessarily with the same selector, or similar classes, i.e., classes defined by the same type (e.g. tuple), are defined at several levels in the class hierarchy. In this case the inheritance property is used so that subclasses inherit methods associated to a super class (the methods may also be redefined at the lower level so as to add some specifics, for example) or subclasses inherit structural properties of a super class. One issue is to assess whether the general methods/classes or the specific methods/classes are defined first. In the former case, we consider that a simple plan is defined in a top-down manner whereas, in the latter case, it is defined in a bottom-up manner.

Second, we analysed more specifically the code reuse activity. In this activity, referred to, in a previous paper (Détienne, 1991), as "new code reuse", the subjects reuse code of the program in progress to develop the code for similar methods or classes. We have analysed the nature of the source in this reuse activity, whether it is a general class or a sibling class, and a general method or a sibling method.

**General-First versus Specific-First for Methods and Classes Definition**

Object-oriented programs are characterised by the fact that similar methods or classes may be defined at several hierarchical levels. Functionally similar methods may be defined in separate classes, in both a parent class and its children, or just in the parent. Both structural and functional properties (data and methods) may be inherited. One issue is to assess, in cases where similar methods or classes may be defined at different levels in the class hierarchy, whether the general methods/classes or the specific methods/classes are defined first by the subjects.

We have analysed the early phase of design, i.e., when the subjects first develop the class model, either on paper or with the editor, before they write methods bodies. This analysis was conducted only for the procedural problem because the inheritance property was mostly used for solving this problem.

Whatever their experience in OOP, all subjects defined the general classes first then the specific classes. A typical example is to define a super class "document" then three subclasses "book", "proceedings" and "journal".

As concerns methods definition at different hierarchical levels, the results are different according to the subjects experience with OOP. The three following situations have been observed:

- from general to specific: A method is first defined at the super class level then similar methods, i.e., methods performing the same function, are defined at the subclasses levels. This was observed for two programmers experienced in OOP and for two beginners in OOP.

-general only: A method is defined only at the super class level. It will be inherited by the subclasses. This was observed for three programmers experienced in OOP and only for one beginner in OOP.

-specific only: A method is defined only at the subclasses level and is not defined at all at the super class level. This was observed for one beginner in OOP and for no programmers experienced in OOP.

Figure 11 present an overview of these results.

## Figure 11 about here

So the programmers experienced in OOP tend to define their methods either at the general level only or at the general level first then at a more specific level. This result shows that they tend to define simple plans in a top-down manner. The results for the beginners in OOP are less clear. In three observations, methods are defined in a top-down manner. However, errors are produced. In one observation, methods are defined only at the specific level. This result does not confirm our hypothesis4 concerning the top-down versus bottom-up manner for developing simple plans.

However, an interesting result is that programmers experienced in OOP tend to develop their methods at only the general level more frequently than beginners in OOP. This result will be discussed in the last section.

We observed that the beginners were not sure about the level at which methods should be defined. Their solutions were incorrect in two cases. For example, when they had several methods which performed the same functionality in different classes, they tried to use the inheritance properties. This is possible when the signature of the method, i.e., its selector and arguments, is the same for the method associated to the super class as it is for the methods associated to the subclasses. However, errors were produced as they generalised this structure without taking into account the signature of methods.

**General-First versus Specific-First in Code Reuse**

We have analysed more specifically the code reuse activity. In this activity, referred to in a previous paper (Détienne, 1991), as "new code reuse" so as to distinguish this activity from "old code reuse", the subjects copy and modify pieces of code for developing the code for similar methods or classes in their program. We have analysed the nature of the source in this reuse activity whether it is a parent class or a sibling class, and a parent method or a sibling method.

A "new code reuse situation" can be described as follows. Very early in the design activity, the programmers judge that different simple plans (different parts of complex plans), elaborated at an abstract level, are instances of the same schema. They make explicit, through verbalisation, that several plans to be developed are exemplars of very common programming algorithms, referred to by the goal they achieve, such as "print values" "search objects" "initialise values of objects", "access to structure and return the field". One of the plans is chosen as the one to be developed first and is thus given the status of "source". Other plans are chosen to be the ones to be developed by copying and modifying the source code, and are given the status of "target".

The subject reasons from a schema and refines it in different ways for elaborating source-solution and target-solution(s). For more clarity, a reuse episode has been defined as a set of behavioural patterns involving the manipulation of one source and one or several targets. This can be interrupted by other activities. As far as the development of different targets is made by reusing the same source, these mechanisms are considered as being part of the same reuse episode. During the source-solution elaboration, some mechanisms have an anticipatory function, i.e., they allow the anticipation of the changes to be handled in elaborating the target-solution(s) from more or less detailed representations of the source-solution (cf. previous paper for a detailed analysis).

We observed 47 reuse episodes (only 1 reuse for syntactic format) for programmers experienced in OOP and, 19 reuse episodes for beginners in OOP (only 3 reuse for syntactic format). In some cases, the writing of methods or classes was reorganised so as to define similar methods or classes in a row, i.e., consecutively. 41 reuses in a row were performed by programmers experienced in OOP and 10 by beginners in OOP.

All experienced programmers, whether or not they developed instances of a schema in a row, developed the most general solutions. When solutions, corresponding to instances of the same schema, had to be coded at several hierarchical levels, the subjects developed the method associated to the most general class first or the most general class first and, after that, the target-solutions corresponding to methods associated to subclasses or corresponding to subclasses.

Rather than using the most general method or class as a source, the beginners in OOP tend to use a sibling entity as the source even when the code of a general class/method, already written, could have been reused. In 6 episodes, for the procedural problem, a hierarchically superior level could be used as source. We observed that:

-for class reuse, the parent class is already defined, however the programmers reuse the code of a sibling subclass for writing the code of other subclasses in 2 episodes.

-for method reuse, we observed (1) the reuse of a sibling method even though the code of a similar method code associated to the parent class, defined before, could have been reused in 2 episodes, and (2) the reuse of a sibling method in 2 episodes, while the method associated to the parent class is defined later on.

Whereas the programmers experienced in OOP tend to develop the code of the general classes/methods first, the beginners in OOP tend to develop the code of the general classes/methods first or develop the code of the specific classes/methods first then develop the code of the general class/method later on. In terms of the development of a simple plan at different levels in the class hierarchy, these results indicate that with a little OOP experience, a simple plan is developed either in a bottom-up manner or in a top-down manner, and that, with more OOP experience, a simple plan is developed in a top-down manner. These results do not give full support to hypothesis4: simple plans to be developed in a top-down manner by programmers who are experienced in OOP and in a bottom-up manner by programmers who are beginners in OOP.

Furthermore, whereas the programmers experienced in OOP use the general entity as a source for more specific entities, the beginners in OOP tend to use a sibling entity as a source even though a general entity was available in several cases.

### 3.3. Revising a Complex Plan

We have analysed the extent to which the complex plans developed in an early design phase are revised by subjects later on. Furthermore, we have analysed, in the final programs, errors which are related to a complex plan.

**Modifying a Complex Plan**

Evaluation mechanisms, such as compiling the programs, lead to revising the solution in progress. In particular, while refining their solutions, all subjects tend to modify the abstract solution developed beforehand. Developing the procedural part in more detail makes them modify the class model. When they refine the functions by writing the code of methods, programmers are able to refine some characteristics of the structure of objects already defined at higher levels of abstraction, e.g., they add attributes in a class or add parameters in a method signature. However they also make more drastic changes to the declarative model. They add new classes and new methods, and move methods from one class to another.

Figure 12a and Figure 12b present the number of modifications made in the models of classes while the subjects developed the procedural parts of their programs.

## Figure 12a and Figure 12b about here

Not surprisingly, more modifications are made by beginners in OOP than by programmers experienced in OOP. Whereas this difference is not significant when we consider the total number of modification (F(1/6)=3.015, P<0.13), it is significant when we consider only the methods modifications (F(1/6)=6.187, p<0.04), e.g., adding/modifying/removing a method and moving a method from one class to another. So, these modifications bear on the definition of a complex plan and the association of parts of this plan with objects. This result bring support to our hypothesis3a: more revisions of the decomposition of classes by beginners in OOP than by programmers experienced in OOP

There is neither a significant effect of the problem type nor a significant interaction effect.

This result exhibits the difficulties beginners in OOP have in articulating the characteristics of objects with the characteristics of procedures. The lack of appropriate knowledge constructed in OOP and the transfer of plans related to procedural languages does not allow the construction of a correct complex plan. When they gain knowledge of the procedure they develop in the procedural part, they revise the static aspects of the global solution, i.e., parts composing a complex plan or the association between the actions of this plan with the objects.

**Errors Related to a Complex Plan**

The final solutions were evaluated by two experts in OOP. This analysis reveals that, even if beginners in OOP tend to revise their complex plans more frequently than programmers experienced in OOP, their final programs still contain many errors (36 errors) concerning the association between the actions of a complex plan with the classes and the structure of the classes. No such errors were found in programs produced by programmers experienced in OOP. Our data bring support to our hypothesis3b: more errors in the decomposition of classes by programmers who are beginners in OOP than by programmers who are experienced in OOP.

In the programs of beginners in OOP, the evaluators found 19 errors corresponding to methods misplaced in the structure of classes and to missing methods for an action developed in the program. It should be remarked that 15 of these errors were found for the procedural problem.

We observed that a common misconception on the part of beginners in OOP is to assimilate the concept of class to the concept of set. Whereas a class represents a family whose object structure is identical, beginners in OOP tend to conceive a class as a set of objects. This misunderstanding causes errors. For example, when creating a method M (for example, search) which processes a set of objects of class A (for example, Book), the subjects tend to associate the method to the class A instead of associating M to a class A' whose type would be set(A) (for example, Books: set(Book)).

In the programs of beginners in OOP, the evaluators found 17 errors corresponding to the structure of classes, e.g., unnecessary classes/objects or using a relational approach for defining classes' characteristics and relationships between classes instead of the is-part-of relationship. It should be remarked that 15 of these errors were found for the declarative problem. These results will be discussed in the next section.

**4. DISCUSSION**

### 4.1. Retrieval of Schemas Related to Different Languages

**Development of Complex Plans**

As regards the development of elements of complex plans attached to different objects, our results support hypothesis 1a for the declarative problem but not for the procedural problem. Programmers who are beginners in OOP tend to develop the actions and the objects separately in their early design, particularly for the declarative problem. For the programmers experienced in OOP, the descriptions of objects and the descriptions of actions tend to be integrated in their first drafts of solutions whichever the problem type. This last result is similar to results found in Kim and Lerch's study (1992). The observed OO designer started his design by identifying the objects and methods of objects at an abstract level.

Our results do not support hypothesis2: complex plans to be developed in a more breadth-first manner by programmers who are experienced in OOP and in a more depth-first manner by programmers who are beginners in OOP.

We found that most of the first drafts of solutions were structured around the objects, whatever the experience of subjects in OOP. This declaratively-driven strategy allows the construction of solutions whose structure is synctatically correct with the OOP language used. Hypothesis1b was not supported by these data; complex plans to be developed on the basis of actions representation by programmers who are beginners in OOP but experienced in procedural languages and on the basis of objects representation by programmers who are experienced in OOP.

An interpretation of the use of this declaratively-driven strategy by the beginners in OOP as well as by the programmers experienced in OOP could be that, whatever their knowledge in OOP, subjects try to construct their first drafts of a solution with a structure which conforms to the structure of OOP programs.

By analysing the order in which actions are generated, we found that, for the programmers experienced in OOP, methods were grouped together mainly by membership to the same class. This result reveals the use of knowledge in which objects and actions are integrated and in which actions are organised around objects. We observed that, for beginners in OOP, methods were grouped together mainly by functional similarity and by execution order. This reflects the use of knowledge in which representation of actions and information on the execution order of actions are prevalent. Hypothesis1c was supported by the data: actions of complex plans to be developed in their execution order more frequently by programmers who are beginners in OOP but experienced in procedural languages than by programmers who are experienced in OOP.

Furthermore, with only a little experience in OOP we observed that complex plans are revised more often and more error are produed. This bring support to hypothesis 3a hypothesis 3b.

Our results suggest that different schemas, depending on subjects' language experience, are used for planning their activity: either schemas related to procedural languages grouping actions in an execution order or schemas related to OO languages integrating actions and objects and with actions organised around objects. The latter type of schemas may be developed through practice with OOP languages and is more adapted to the constraints which must be taken into account in designing with this kind of languages, e.g., making explicit the links between objects and actions.

**Development of Simple Plans**

As concerns the development of simple plans at different levels in the class hierarchy, our results indicate that with a little OOP experience, simple plans are developed either in a top-down manner or in a bottom-up manner. The beginners in OOP tend to develop the definition of simple plans at the general level first. However, errors were produced. Only in one observation are methods defined only at the specific level. The beginners in OOP tend to develop the code of simple plans either in a top-down manner by developing the code of the general classes/methods first or in a bottom-up manner by developing the code of the specific classes/methods first and then developing the code of the general class/method later on. They tend to use a sibling entity as a source even when a general entity is available.

With more OOP experience, our results show that a simple plan is developed in a strictly top-down manner either by developing the definition of a simple plan at a general level first before developing it at a more specific level, or by developing the code of a general entity first then reusing it for developing the code of more specific entities. So we found that, at least for the definition of methods, the beginners as well as the programmers experienced in OOP tend to adopt a top-down approach. This does not confirm hypothesis4 which was that programmers who are beginners in OOP would follow a bottom-up approach whereas programmers experienced in OOP would follow a top-down approach. An explanation could be that, for the development of simple plans, even the beginners would possess the knowledge necessary.

However, an interesting result is that programmers experienced in OOP tend to develop their methods at the general level only whereas beginners in OOP tend to develop them at a general level and also, often in an incorrect way, at a more specific level. This suggests that, even if a simple schema is available for beginners in OOP, the knowledge about the validity of associating a simple plan at the correct hierarchical level is not available for these subjects.

It should be remarked that the results on code reuse are different from those found in a single-subject study of a software developer working in an OOP environment conducted by Lange and Moher (1989). These authors found that the designer, who was experienced in OOP, chose first a sibling entity as a source. If a sibling entity did not exist then the parent class was chosen as the source. This result was found, in our study, for the beginners in OOP but not for the programmers experienced in OOP. It is possible that the choice of a reuse strategy may depend, for programmers experienced in OOP, on the size of the program under development (the program in the Lange and Moher's study was larger than the programs in our study) and also on the amount of modifications the subjects think they have to make according to the chosen source.

Furthermore, it should be remarked that even though functional similarity did not seem to play an important role in the definition of methods in an early phase of design by programmers experienced in OOP, this relationship was important in organising code generation of methods. In code generation, the programmers experienced in OOP, as well as the beginners in OOP, organise the generation of code of functionally similar methods in a row so as to facilitate code reuse.

**Implications**

The results of this study highlight the diversity of strategies used by programmers developing programs with an OOP language. Support systems should take these various strategies into account. Some strategies are dependent on previous experience in procedural languages. The transfer of knowledge acquired with procedural languages have been shown to affect the kind of design strategy used and the product of the design activity, i.e., the accuracy of the solutions produced.

Programmers who are beginners in OOP but experienced in procedural languages tend to use strategies based on a dynamic view of the solution. We have observed that the functional characteristics of the solution and the execution order of actions are important in the design activity of these subjects. The design methods and the design tools should support this kind of strategy.

The transfer of knowledge acquired with procedural languages affects the accuracy of the solutions produced. Transfer mechanisms of schemas related to procedural languages have been shown to produce errors whereas strategies are used in order to avoid making errors. Our study show that the implementation, in an unfamiliar language, of transferred programming schemas may cause the production of errors related to the complex plan of the solution.

We have shown in a previous paper (Détienne, 1993) that programmers use design strategies in order to compensate for lack of knowledge about how to use the programming plans. Two strategies were observed: (1) using exemplars of solutions provided by an example-program written in the unfamiliar language so as to infer the rules of validity of particular plans and, (2) using weak problem solving methods such as generate-test-and-debug.

Furthermore, in the present study, we observed more revisions of the decomposition of classes by programmers beginners in OOP than by programmers experienced in OOP. Scholtz and Wiedenbeck (1991) observed that failing to implement transferred schemas at the code level may lead to plan revision at more abstract levels: there are iterations between the implementation level and the solution elaboration levels.

Tools should support the transfer mechanisms, in particular, by providing feedback about the validity of the transferred schemas at levels of abstraction higher than the implementation level. Training curricula should also take into account the previous knowledge experienced programmers may transfer when acquiring a new language.

**4.2. Effect of Problem Characteristics**

According to hypothesis5, we expected the design activity to be easier for a declarative problem than for a procedural problem, at least for the programmers who are beginners in OOP. This could be measured in terms of strategies and errors.

Previous studies have shown effects of problem characteristics on the design strategies. Two kinds of design problems were used in this experiment: a "declarative" problem and a "procedural" problem. The distinction between these two types of problems was made on the basis of categories formed by programmers experienced in procedural programming and on the basis of their design activity (Hoc, 1983; 1988). It was shown that, for declarative problems, the data structure is strong and guides program development and that, for procedural problems, the structure of the procedure is strong and guides program development.

Our results showed that the type of problem had no effect on the strategies followed by the subjects except for the integration versus no integration between the description of objects and the description of actions. In an early design phase, it was found that, for the beginners in OOP, the object and action descriptions tended not to be integrated for the declarative problem whereas they were integrated for the procedural problem.

Our results showed that, for beginners in OOP, the number of errors related to a complex plan was almost the same whichever the problem type. This result does not confirm hypothesis 5.

However, one interesting result was that the type of errors produced by the beginners in OOP was related to the type of problem. Errors corresponding to the structure of classes, e.g., unnecessary classes/objects, were produced mainly for the declarative problem. Errors corresponding to methods being misplaced in the structure of classes and to missing methods for an action developed in the program, were produced mainly for the procedural problem.

In the paradigm of object-oriented programming, identifying classes and their characteristics is important in the design process. In a declarative problem, we expected these aspects to be more straightforward to analyse as the data structure is strong. However. our results tend to show that the definition of the structure of classes is error prone mostly for this problem. This suggests that when a strong data structure known for procedural languages may be transferred, the resulting representation includes unnecessary classes and/or objects structured in a way more appropriate to procedural programming. For the procedural problem, this suggests that when a strong structure of procedure known for procedural languages may be transferred, the structure of this procedure is developed as a complex plan and linking parts of this plan with classes is error prone.

**4.3. Limitations of the Study and Future Studies**

To conclude, we should point out several limitations of this study. First, even though this study provides interesting insight into the design strategies and knowledge used in the design activity with an object-oriented language, it is necessary to conduct further studies with a greater number of subjects and with less simple problems so as to generalise our results.

Second, the beginners in this study were observed at the very beginning of their use of an OOP language for solving a problem. An issue is how long the differences observed between beginners and experienced programmers persist as beginners get more experienced with the OOP language. Another related issue is how the strategies used may be determined by the difficulty of the problem. In the present study, the problems were relatively simple compared to real OOP problems. We could assume that, for difficult problems or subproblems, even the experienced programmers could behave as beginners, using strategies based on the execution order of methods/actions in the program. These issues are particularly important for developing support to the design activity which would take into account which conditions, external and internal to the subjects, trigger a particular strategy.

Third, whereas the use of design strategies may reveal the subject's knowledge, they also reflect other characteristics of the task such as environment characteristics and language characteristics. We feel that complementary analyses, conducted with other tasks could be carried out in order to generalise our findings and also to provide complementary results. Tasks such as code grouping or problem categorisation (Détienne, Borne & Chatel, 1993) are potentially of interest in order to achieve this aim. In the grouping tasks, programs have to be segmented in order to group lines together. Subjects are asked to make explicit the criteria used for this grouping. This provides information about which chunks of code in programs are related to single programming plans. Problem categories provided by experts in a categorisation task give information about problem dimensions related to the design strategies used by experts.

Finally, other OO languages should be used to generalise these results. The $CO_2$ language has procedural characteristics which are not present in other OO languages. This may affect the strategies used.

*Published in Human-Computer Interaction, 1995, 10 (2 & 3), 129-170.*
*http://hci-journal.com/*

**NOTES**

*Background.* This article is partly based on two earlier preliminary papers which focused on code reuse and several characteristics of strategies, specifically evaluation strategies, used with an OOP language (Détienne, 1991; 1993); complementary analyses have been conducted so as to examine more specifically the knowledge used in designing with an OOP language.

*Acknowledgements.* We would like to express our thanks to Pascal Claudel for collecting the protocols, to Philippe Delsol for participating in one part of the protocols analysis, to Sophie Gamerman for her help in the realisation of this experiment and to Willemien Visser for her comments on an earlier draft of this paper. Special thanks go to Robert Rist for his constructive comments given during informal discussions on this study.

*Support.* This study was conducted under a research contract with the GIP Altaïr (Altaïr is a consortium funded by IN2, INRIA and LRI University Paris XI).

*Author's Address.* Françoise Détienne, Projet de Psychologie Ergonomique, INRIA (Institut National de Recherche en Informatique et Automatique), Domaine de Voluceau, Rocquencourt, BP 105, 78153 Le Chesnay, cedex, France. Email: detienne@nuri.inria.fr

**REFERENCES**

Adelson, B. & Soloway, E.(1985) The role of domain experience in Software Design. *IEEE Transactions on Software Engineering*, *11 (11)*, 1351-1360.

Black, J. B., Kay, D. S. & Soloway, E. (1986) Goal and Plan Knowledge Representations: From Stories to Text Editors and Programs. In J.M. Carroll (Ed.), *Interfacing Thought: Cognitive aspects of human-computer interaction*. Cambridge, Mass: MIT Press.

Booch, G. (1991) *Object-Oriented Design with Applications*. Benjamin/Cummings, Redwood City, CA.

Davies, S. (1991) The Role of Notation and Knowledge Representation in the Determination of Programming Strategy: A Framework for Integrating Models of Programming Behavior. *Cognitive Science*, *15*, 547-572.

Détienne, F.(1990a) Expert Programming Knowledge: A Schema-Based Approach. In J-M. Hoc, T.R.G. Green, R. Samurçay, D. Gilmore (Eds). *Psychology of programming* (pp 205-222). People and Computer Series, Academic Press.

Détienne, F. (1990b) Program Understanding and Knowledge Organization: the Influence of Acquired Schemas. In P. Falzon (Ed). *Cognitive Ergonomics: understanding, learning and designing Human-Computer Interaction* (pp 245-256). London: Academic Press.

Détienne, F.(1991) Reasoning from a schema and from an analog in software code reuse. In J. Koenemann-Belliveau, T. Moher & S. P. Robertson (Eds). *Empirical studies of programmers, Fourth Workshop* (pp 5-22). Norwood, NJ: Ablex.

Détienne, F. (1993) Acquiring experience in object-oriented programming: effects on design strategies. In E. Lemut, B. du Boulay and G. Dettori (Eds). *Cognitive Models and Intelligent Environments for Learning Programming.* Springer-Verlag. NATO ASI Series.

Détienne, F., Borne, I. & Chatel, S. (1993) The activity of Design with Object-Oriented Languages. *Proceedings of the INTERCHI' 93 Research Symposium*. Amsterdam, NL, April 23-24.

Guindon, R.(1990) Knowledge exploited by experts during software system design. *International Journal of Man-Machine Studies, 33(3)*, 241-360.

Hoc, J-M.(1988) Towards effective computer aids to planning in computer programming: theoretical concern and empirical evidence drawn from assessment of a prototype. In G.C. van der Veer, T.R.G. Green, J-M. Hoc & D. Murray (Eds): *Working with computers, theory versus outcome* s(pp 215-247), London, Academic press.

Hoc, J-M. (1983) Une méthode de classification préalable des problèmes d'un domaine pour l'analyse des stratégies de résolution: la programmation informatique chez des professionnels. *Le Travail Humain, 46 (3)*, pp. 205-217.

Kay, D.S. & Black, J.B. (1985) The evolution of knowledge representations with increasing expertise in using systems. *Proceedings of the Seventh annual meeting of the cognitive science society*. Irvine, CA.

Kim, J. & Lerch, J. (1992) Towards a Model of Cognitive Process in Logical Design: Comparing Object-Oriented and Traditionnal Functional Decomposition Software Methodologies. *Proceedings of CHI'92 Conference on Human Factors in Computing Systems*, 489-498, Addison-Wesley Publ: ACM Press .

Lange, B.M. & Moher, T.G. (1989) Some Strategies of Reuse in an Object-Oriented Programming Environment. *Proceedings of CHI'89*, 69-73.

Lecluse, C. and Richard, P. (1989) The O2 Database Programming Language. *Proceedings of International Conference on Very Large Data Bases*, Amsterdam, August 26, 1989.

O.Deux et al.(1989) *The Story of O2* (Tech. Rep. 37-89). GIP Altaïr, Rocquencourt.

Rist, R. (1986) Plans in Programming: Definition, Demonstration, and Development. In E. Soloway & S. Iyengar (Eds.). *Empirical Studies of Programmers: First Workshop* (pp 28-47). Norwood, N.J.: Ablex Publishing Corporation.

Rist, R. (1991) Knowledge creation and retrieval in program design: a comparison of novice and intermediate student programmers. *Human-Computer Interaction, 6*, 1-46.

Rist, R. (1994) Search through Multiple Representations: In D. Gilmore, R. Winder and F. Détienne (Eds), *User-Centred Requirements for Software Engineering Environments*. (pp. 165-176). Berlin: Springer-Verlag NATO ASI Series.

Rist, R. and Terwilliger, R. (1994) *Object-Oriented Progrmming in Eiffel*. Sydney: Prentice-Hall.

Robertson, S.P. and Yu, C. C. (1990) Common cognitive representations of program code across tasks and languages. *International Journal of Man-Machine Studies, 33*, 343-360.

Rosson, M. B. and Alpert, S. R. (1988) *The Cognitive Consequences of Object-Oriented Design* (Res. Rep. RC 14191), IBM, N.Y.

Scholtz, J. and Wiedenbeck, S. (1990) Learning to Program in Another Language. *Proceedings of INTERACT'90*, 925-930, North Holland, p 925-930.

Scholtz, J. and Wiedenbeck, S. (1992) Using an Unfamiliar Programming Language. In A. Monk, D. Diaper & M.D. Harrison (Eds). *People and Computers VII.* Cambridge: Cambridge University Press.

Soloway, E., Ehrlich, K. & Bonar, J. (1982) Tapping into Tacit Programming Knowledge. *Human Factors in Computer Systems, 15-17*, 52-57.

Soloway, E. & Ehrlich, K.(1984) Empirical Studies of Programming Knowledge. *IEEE Transactions on Software Engineering, SE-10 (5)*, 595-609.

Wirfs-Brock, R.B. & Johnson, E.E. (1990) Surveying Current Research in Object-Oriented Design. Communications of the ACM, 33 (9), 104-123.

**APPENDIX A: Profile of subjects, i.e., programming languages which are frequently used or known but not used frequently by each subject**

** : frequently used language

* : known language which is not used frequently

<blank>: not known language

| Type of language | Programming language | Beginners in OOP | | | | Experienced in OOP | | | |
|---|---|---|---|---|---|---|---|---|---|
| | | B1 | B2 | B3 | B4 | E1 | E2 | E3 | E4 |
| Procedural | C | ** | ** | ** | ** | ** | ** | ** | ** |
| | Pascal | | ** | ** | * | ** | * | * | * |
| | Fortran | | ** | * | | * | * | | |
| | Basic | | | * | * | ** | | | |
| | Cobol | ** | | | | | | | |
| Functional | Lisp | | | | * | | | ** | ** |
| Declarative | Prolog | | | | | | | ** | |
| Hybrid | Ada | | | | * | | | | |
| Other | Assembler | | | * | * | | * | | |
| Object-Oriented | CO2 | | | | | ** | ** | ** | ** |
| | Smalltalk | | | | | * | | | |

**APPENDIX B: Problem statements**

**APPENDIX B1: Procedural Problem Statement, Library Management Problem**

A library manager wishes to have an automatic documentation retrieval program for all the library books. Each book is described by several key-words. One wants to obtain books on the basis of two types of questions:

-books associated to one key-word

-books associated to two key-words

With the data base that you create, you should obtain the following results. For a given question, print:

-the question comprising one key-word or two key-words separated by "and"

-the answer which can be of three distinct types:

    -a list of books, if this list is not empty.

    For every book, the title and the year is displayed. Furthermore, for the proceedings, the location of the conference is displayed, and for the journals, the volume is displayed.

    -the response "no books of this type" if this is the case

    -for each invalid key-word: the key-word followed by "not registered"

**APPENDIX B2: Declarative Problem Statement, Financial Management Problem**

In an aeronautic club, the manager wishes to know at the end of the year the amounts to charge the members, the club's annual turn-over and the flight statistics. One has the diverse rates and information about the members and the flights. The rates depend on the members' situations and on some characteristics of the flights:

The rules for calculating the rates are as follows:

-two types of subscription reductions can be added: 25% for a member younger than 20 and, 5% per child of the member's family;

-the insurance premium is fixed;

-each plane has its own rate per hour and flights costs are calculated by the hour.

With the data base that you create, you should obtain the following results:

-for each member: identification number, family name, detailed amount to be paid (subscription, insurance premium and total amount to pay for the hours of flight), total amount to be paid;

-club's annual turn-over

Remark:

There is a predefined class "Person" which can be used in this solution. The selectors you can use are:

-the "family-name" selector which returns a string;

-the "age" selector which returns an integer;

**APPENDIX C: Characteristics of the programs produced by the subjects**

*an embedding degree equal to 1 corresponds to "no embedding"

**a depth equal to 1 corresponds to a flat structure

**APPENDIX C1: Characteristics of the programs produced for the procedural problem**

|  | Beginners in OOP | | | | Experienced in OOP | | | |
|---|---|---|---|---|---|---|---|---|
|  | B1 | B2 | B3 | B4 | E1 | E2 | E3 | E4 |
| Number of defined classes | 5 | 4 | 5 | 4 | 7 | 8 | 4 | 4 |
| Number of defined methods | 10 | 9 | 13 | 10 | 7 | 21 | 21 | 8 |
| Number of coded methods | 0 | 9 | 12 | 10 | 1 | 21 | 21 | 8 |
| use of the is-a relationship (inherits) | 2 | 2 | 3 | 2 | 2 | 2 | 2 | 2 |
| use of the is-part-of relationship | 1 | 1 | 0 | 1 | 4 | 7 | 1 | 0 |
| Maximum embedding degree in composite classes* | 2 | 2 | 1 | 2 | 3 | 2 | 2 | 1 |
| Maximum depth of inheritance** | 2 | 2 | 2 | 2 | 2 | 2 | 2 | 2 |

**APPENDIX C2: Characteristics of the programs produced for the declarative problem**

|  | Beginners in OOP | | | | Experienced in OOP | | | |
|---|---|---|---|---|---|---|---|---|
|  | B1 | B2 | B3 | B4 | E1 | E2 | E3 | E4 |
| Number of defined classes | 7 | 5 | 5 | 3 | 7 | 4 | 4 | 4 |
| Number of defined methods | 27 | 25 | 16 | 4 | 14 | 14 | 12 | 11 |
| Number of coded methods | 18 | 10 | 10 | 4 | 13 | 12 | 11 | 6 |
| use of the is-a relationship (inherits) | 0 | 1 | 1 | 1 | 1 | 0 | 1 | 0 |
| use of the is-part-of relationship | 3 | 3 | 3 | 1 | 3 | 4 | 2 | 4 |
| Maximum embedding degree in composite classes* | 2 | 2 | 3 | 2 | 3 | 4 | 3 | 3 |
| Maximum degree of classes hierarchisation** | 1 | 2 | 2 | 2 | 2 | 1 | 2 | 1 |

**APPENDIX C3: Characteristics of the programs: average by subject type**

|  | Beginners in OOP | Experienced in OOP |
|---|---|---|
| Number of defined classes | 4.75 | 5.25 |
| Number of defined methods | 14.25 | 13.5 |
| Number of coded methods | 9.13 | 11.63 |
| use of the is-a relationship (inherits) | 1.5 | 1.25 |
| use of the is-part-of relationship | 1.63 | 3.13 |
| Maximum embedding degree in composite classes* | 2 | 2.63 |
| Maximum degree of classes hierarchisation** | 1.88 | 1.75 |
| Number of instructions | 103.5 | 184.5 |

---

[1] Many research are conducted on object-oriented design (Wirfs-Brock & Johnson, 1990) but very few studies are empirical.

[2] For example, for a problem of stock management, the typical data structure is a record (name of file, descriptor of file, etc.) and the functions which are typical of this problem are allocation (creation or insertion), destruction, and search.

[3] Evaluation strategies used by subjects according to their experience in OOP have been discussed in a previous paper (Détienne, 1993).

[4] The author was the primary scorer.

*Figure 1a*. **Example of a declarative part.**

*Figure1b*. **Example of a procedural part.**

*Figure 2a*. **Characteristics of the first drafts of solutions produced by programmers beginners in OOP.**

*Figure 2b*. **Characteristics of the first drafts of solutions produced by programmers experienced in OOP.**

*Figure 3a*. **Integration versus no integration between the description of actions and the description of objects by subject type for the procedural problem.**

*Figure 3b*. **Integration versus no integration between the description of actions and the description of objects by subject type for the declarative problem.**

*Figure 4*. **Excerpt of B2's protocol for the declarative problem.**

*Figure 5*. **Excerpt of E1's protocol for the procedural problem.**

*Figure 6*. **Declaratively-driven strategy versus procedurally-driven strategy: by subject type and by problem type.**

*Figure 7*..**Excerpt of E4's protocol for the procedural problem.**

*Figure 8*. **Excerpt of B1's protocol for the procedural problem.**

*Figure 9*. **Number and percentage of chunks of methods according to the relationship between methods.**

*Figure 10*. **Number of methods defined in the class model before developing the procedural part.**

*Figure 11*. **levels at which methods are defined; number of times a given situation has been observed by subject.**

*Figure 12a*. **Total number of modifications made in the class model.**

*Figure 12b*. **Number of method modifications made in the class model.**

*Figure 1a*. **Example of a declarative part.**

```
add class Book
    type tuple (title:string,
    year: integer)
    method title: string
    method show
...
add class Proceedings inherits Book
    type tuple (place: string)
    method show
...
add class Journal inherits Book
    type tuple (vol: integer)
    method show
...
```

*Figure1b.* **Example of a procedural part.**

```
body title: string in class Book
{return (self->title);}
...
body show in class Book
{printf("title=%s, year=%d", self->title, self->year);}
....
```

*Figure 2a.* **Characteristics of the first drafts of solutions produced by programmers beginners in OOP.** For each number , we describe the notes taken in a single draft: 1 for the first draft, 2 for the second draft and so on. Descriptions may be:

-"list of objects (as classes)": the notes can be described as a single list of objects defined or not defined as classes

-"list of actions (as methods)": the notes can be described as a single list of actions defined or not defined as methods

-"list of actions as methods; for each action, several objects are associated": the notes can be described as a list of actions with objects associated to actions.

-"list of objects as classes; for each object, several actions are associated": the notes can be described as a list of objects with actions associated to objects.

|  | B1 | B2 | B3 | B4 |
|---|---|---|---|---|
| procedural problem | 1. list of objects as classes<br>2. list of actions as methods; for each action, several objects are associated<br>3. list of objects as classes; for each object, several actions are associated<br>4. list of actions; for each action, several objects are associated | 1. list of objects as classes<br>2. list of objects as classes; for each object, several actions are associated | 1. list of objects as classes; for each object, several actions are associated | 1. list of objects as classes; for each object, several actions are associated |
| declarative problem | 1. list of objects<br>2. list of actions + *simulation*<br>3. list of objects as classes<br>4. list of actions as methods; for each action, several objects are associated | 1. list of objects as classes<br>2. list of actions + *simulation*<br>3. list of objects as classes; for each object, several actions are associated | 1. list of objects as classes<br>2. list actions as methods; for each action, several objects are associated | 1. list of objects as classes; for each object, several actions are associated |

*Figure 2b*. **Characteristics of the first drafts of solutions produced by programmers experienced in OOP.**

|  | E1 | E2 | E3 | E4 |
|---|---|---|---|---|
| Procedural problem | 1. list of objects as classes<br>2. list of objects as classes<br>3. list of actions as methods; for each action, several objects are associated<br>+ *simulation* | 1. list of objects as classes<br>2. list of objects as classes; for each object, several actions are associated<br>+ *simulation*<br>3. list of objects as classes; for each object, several actions are associated (alternative solution) | 1. list of objects as classes; for each object, several actions are associated<br>2. list of objects as classes; for each object, several actions are associated | 1. list of objects as classes<br>2. list of objects as classes; for each object, several actions are associated |
| Declarative problem | 1. list of objects as classes; for each object, several actions are associated<br>2. list of objects as classes; for each object, several actions are associated | 1. list of objects as classes; for each object, several actions are associated<br>+ *simulation*<br>2. list of objects as classes; for each object, several actions are associated | 1. list of objects as classes; for each object, several actions are associated | 1. list of objects as classes<br>2. list objects as classes; for each object, several actions are associated |

*Figure 3a.* **Integration versus no integration between the description of actions and the description of objects by subject type for the procedural problem.**

|                     | Integration | No integration |
|---------------------|-------------|----------------|
| Experienced in OOP  | 3           | 1              |
| Beginners in OOP    | 3           | 1              |

*Figure 3b.* **Integration versus no integration between the description of actions and the description of objects by subject type for the declarative problem.**

|                     | Integration | No integration |
|---------------------|-------------|----------------|
| Experienced in OOP  | 4           | 0              |
| Beginners in OOP    | 1           | 3              |

*Figure 4.* **Excerpt of B2's protocol for the declarative problem.**

```
"Member                                         plane
person                  -name                   -plane identification
                        -age                    -rate per hour
                        -number of children
-identification number                          -duration of flight
-payment of subscription                         during the current
-consumption                                     year

 Flight                                         Club
-member                                         -set of members
-date/start                                     -turn-over
-date/end
```
---

<u>For each flight</u>:      -register the date of flight
                     -calculate the consumption
                     -update member

<u>At the end of the year</u>
the last day
Any member:
-calculate subscription (age-children)
 -display: identification number, name, subscription, premium, consumption, total
  of three sums
- add in the turn-over
-display the turn-over"

*Figure 5.* **Excerpt of E1's protocol for the procedural problem.**

---

<u>Definition of data</u>

.......
<u>Definitive solution</u>
<u>a. Classes</u>
-class Key-words:{string}
-class: **BOOK**
    type tuple (title::string,
        year:integer
        key-words:Key-words
-class Proceedings: sub-class of Book
    type tuple (place::string)
-class Journal: sub-class of Book
    type tuple (volume::integer)
<u>b. Named objects</u>: registered key-words: Key-words
    TheBooks: set{Book}
    TheProceedings:set{Proceedings}
    TheJournals: set{Journal}

<u>Data processing</u>

-prompt for asking a question
    -choice between Q1 and Q2
-display the question
-analyse the question
    check if the key-words are registered
        else display a message
        "the descriptor is not registered" and stop
        if registered continue
-search in the set of books which are available
    -put in a working list each book which has the key-word(s)
if the working list is empty
    display "no book of this type"
else: display the list

---

*Figure 6.* **Declaratively-driven strategy versus procedurally-driven strategy: by subject type and by problem type.**

|  | Beginners in OOP | | | Experienced in OOP | | |
|---|---|---|---|---|---|---|
|  | Procedural Problem | Declarative Problem | Total | Procedural Problem | Declarative Problem | Total |
| Declaratively-driven strategy | 4 | 2 | 6 | 3 | 4 | 7 |
| procedurally-driven strategy | 1 | 2 | 3 | 1 | 0 | 1 |

*Figure 7.*.**Excerpt of E4's protocol for the procedural problem.**

```
" add class Book
type tuple (title: string
year: integer,
key-word: list(string)
)
public * in class book

add method enter: integer in class Book is public
add method print (flag: integer) in class Book is public
add method match (m:string):integer in class Book

add class Proceedings inherits Book
type tuple (location:string)

add class Journal inherits Book
type tuple (volume: integer,
number: integer)

add value Library: set(Book)

add class Dummy type integer
add object Dummy: Dummy
add method Dummy in class Dummy is public"
```

*Figure 8.* **Excerpt of B1's protocol for the procedural problem.**

```
"Methods
     ? key-word of book
     ? Key-word included in list of Key-words
     ? Display   ->Book
                  ->Journal
                   -> Proceedings
     ? init
     ?Create/modify/delete
     ? method for answering question        3 options
"
```

*Figure 9.* **Number and percentage of chunks of methods according to the relationship between methods.**

|  | Membership in the same class | Functional similarity | Execution Order | TOTAL |
|---|---|---|---|---|
| Experienced in OOP | 25 (80.5%) | 5 (16%) | 1 (3.5%) | 31 |
| Beginners in OOP | 4 (22%) | 8 (44.5%) | 6 (33.5%) | 18 |

*Figure 10.* **Number of methods defined in the class model before developing the procedural part.**

|  | B1 | B2 | B3 | B4 | E1 | E2 | E3 | E4 |
|---|---|---|---|---|---|---|---|---|
| procedural problem | 9 | 8 | 8 | 2 | 9 | 11 | 19 | 4 |
| declarative problem | 18 | 5 | 3 | 4 | 12 | 13 | 13 | 9 |

*Figure 11.* **levels at which methods are defined; number of times a given situation has been observed by subject**

|  | B1 | B2 | B3 | B4 | E1 | E2 | E3 | E4 |
|---|---|---|---|---|---|---|---|---|
| from general to specific | 1 | 0 | 2 | 0 | 1 | 0 | 1 | 0 |
| general only | 0 | 0 | 0 | 2 | 0 | 4 | 3 | 3 |
| specific only | 0 | 2 | 0 | 0 | 0 | 0 | 0 | 0 |

*Figure 12a*. **Total number of modifications made in the class model.**

|  | B1 | B2 | B3 | B4 | E1 | E2 | E3 | E4 |
|---|---|---|---|---|---|---|---|---|
| procedural problem | 16 | 22 | 21 | 32 | 22 | 29 | 10 | 9 |
| declarative problem | 7 | 46 | 29 | 15 | 9 | 6 | 23 | 13 |

*Figure 12b.* **Number of method modifications made in the class model.**

|  | B1 | B2 | B3 | B4 | E1 | E2 | E3 | E4 |
|---|---|---|---|---|---|---|---|---|
| procedural problem | 13 | 13 | 18 | 20 | 7 | 12 | 10 | 9 |
| declarative problem | 4 | 38 | 22 | 12 | 4 | 4 | 14 | 3 |